\documentclass{aa}
\usepackage{aas_macros}
\usepackage{epsfig}
\usepackage{rotating}
\usepackage{graphicx}
\newcommand{\fermi}{{\sl Fermi}}
\newcommand{\fermin}{{\sl Fermi Gamma-ray Space Telescope}}

\newcommand{\gron}{{\sl Compton Gamma Ray Observatory}}
\newcommand{\gro}{{\sl CGRO}}
\newcommand{\swift}{{\sl Swift}}
\newcommand{\chan}{{\em Chandra}}
\newcommand{\xmm}{{\em XMM-Newton}}
\newcommand{\suz}{{\em Suzaku}}
\newcommand{\swi}{{\em Swift}}
\newcommand{\vltn}{{\em Very Large Telescope}}
\newcommand{\vlt}{{\em VLT}}

\newcommand{\hst}{{\em HST}}
\newcommand{\fors}{{\em FORS2}}
\newcommand{\forsn}{{\em FOcal Reducer/low dispersion Spectrograph}}

\def \psr{PSR\, J1028$-$5819}

\begin{document}

\title{VLT and \emph{Suzaku} observations of the \emph{Fermi} pulsar \\ PSR\, J1028$-$5819\thanks{Based on observations collected at ESO, Paranal, under Programme 086.D-0392(B)}}

\author{R. P. Mignani\inst{1,2}
\and
M. Razzano\inst{3,4,10}
\and
P. Esposito\inst{5}
\and 
A. De Luca\inst{6,7,8}
\and
M. Marelli\inst{7,9}
\and
S.R. Oates\inst{1}
\and
P. Saz-Parkinson\inst{10}
}

   \institute{Mullard Space Science Laboratory, University College London, Holmbury St. Mary, Dorking, Surrey, RH5 6NT, UK
   \and{Kepler Institute of Astronomy, University of Zielona G\'ora, Lubuska 2, 65-265, Zielona G\'ora, Poland}
   \and{Istituto Nazionale di Fisica Nucleare, Sezione di Pisa, I-56127 Pisa, Italy}
   \and{Dipartimento di Fisica ``E. Fermi'', Universit\`a di Pisa, 56127 Pisa, Italy}
    \and{INAF Ð Osservatorio Astronomico di Cagliari, localit\`a Poggio dei Pini, Strada 54, I-09012 Capoterra, Italy}
   \and{IUSS - Istituto Universitario di Studi Superiori, viale Lungo Ticino Sforza, 56, 27100, Pavia, Italy}
   \and{INAF - Istituto di Astrofisica Spaziale e Fisica Cosmica Milano, via E. Bassini 15, 20133, Milano, Italy}
   \and{INFN - Istituto Nazionale di Fisica Nucleare, sezione di Pavia, via A. Bassi 6, 27100, Pavia, Italy}
   \and{Universit\`a  degli Studi dell'Insubria, Via Ravasi 2, 21100, Varese, Italy}
   \and{Santa Cruz Institute for Particle Physics, University of California, Santa Cruz, CA 95064}
 }

\titlerunning{\vlt\ observations of \psr}

\authorrunning{Mignani et al.}
\offprints{R. P. Mignani; rm2@mssl.ucl.ac.uk}

\date{Received ...; accepted ...}

\abstract{The launch of the \fermin\ in 2008 opened new perspectives in the multi-wavelength studies of neutron stars, with more than 100  $\gamma$-ray pulsars now detected.  While most  \fermi\ pulsars have been already observed in the X-rays with \chan\ and \xmm, optical observations with 8m-class telescopes only exist for a tiny fraction of them. 
} 
{Here, we aim at searching for optical emission from the \fermi\ pulsar PSR\, J1028$-$5819 ($P=91.4$ ms).  With a spin-down age $\tau =92.1$ kyr and a rotational energy loss rate $\dot{E} \sim 8.43 \times 10^{35}$ erg s$^{-1}$, \psr\ can be considered a transition object between the young, Vela-like pulsars and the middle-aged ones. At a distance of $\sim 2.3$ kpc and with a relatively low hydrogen column density, \psr\  is a good potential target for 8m-class telescopes. }   
{Due to its recent discovery, no optical observations of this pulsar have been reported so far. We used optical images  taken with the \vltn\ (\vlt) in the $B$ and $V$ bands  to search for the optical counterpart of \psr\ or constrain its optical brightness. At the same time, we used an archival  \suz\ observation to confirm the preliminary identification of the pulsar's X-ray counterpart obtained by \swi.}
{Due to the large uncertainty on the pulsar's radio position and the presence of a bright ($V=13.2$) early F-type star at $\la 4\arcsec$ (Star A),  we could not detect its counterpart down to flux limits of $B\sim25.4$ and $V\sim25.3$, the deepest obtained so far for \psr.  From the \suz\ observations, we found that the X-ray spectrum of the pulsar's candidate counterpart is best-fit by a power-law with spectral index $\Gamma_{\rm X} = 1.7\pm0.2$ and an absorption column density $N_{\mathrm{H}}<10^{21}$ cm$^{-2}$, which would support the proposed X-ray identification. Moreover, we found possible evidence for the presence of diffuse emission around the pulsar. If real, and associated with a pulsar wind nebula (PWN), its surface brightness and angular extent would be compatible with the expectations for a $\sim$ 100 kyr old pulsar at the \psr\ distance.}
{A far more accurate radio position for \psr\  is necessary to better determine its position relative to Star A. Future high-spatial resolution observations with both the \hst\ and \chan\  are better suited to disentangle the optical emission of \psr\ against the halo of Star A and to confirm the existence of the candidate PWN.}

 
   \maketitle

\section{Introduction}
The launch of NASA's {\em Fermi} Gamma-ray Space Telescope in June 2008  marked a breakthrough in  $\gamma$-ray observations of pulsars.  The {\em Large Area Telescope} ({\em LAT};  Atwood et al.\ 2009) has now detected more than 100  $\gamma$-ray pulsars, 
to be compared with the 7 detected by the  \gron\ (\gro) in 10 years of operation (see, e.g.  Thompson 2008 for a recent review).  Some of them are associated with known, or recently discovered, radio pulsars, while others are yet undetected in the radio despite of deep searches and are, thus, members of the long-hypothesised class of radio-silent $\gamma$-ray pulsars, of which Geminga was the first example (e.g., Bignami \& Caraveo 1996; Abdo et al.\ 2010a).  In both cases, multi-wavelength follow-ups are important to study the pulsar radiation processes over several decades in energy, to constrain the source distance and position, and to search for pulsar wind nebulae (PWNe), still challenging targets in $\gamma$-rays despite of the unprecedented spatial resolution of the \fermi-LAT (e.g., Ackermann et al.\ 2011).  
In the X-rays, many \fermi\ pulsars were already observed prior to their $\gamma$-ray detection, while for the others systematic follow-up observations have been carried out recently with \chan, \xmm, and {\em Swift} (see, e.g. Marelli et al.\ 2011 for a summary of the X-ray observations of \fermi\ pulsars).  In the optical, the source coverage with 8m-class telescopes is still sparse  with only a few \fermi\ pulsars observed beforehand and only a few deep follow-up observations on well-selected targets have been carried out so far  (Mignani et al.\ 2011; Shearer et al.\  2012). 

The radio pulsar PSR\, J1028$-$5819 ($P=91.4$ ms) was discovered by Keith et al.\ (2008) during a 3.1 GHz survey of a sample of low-latitude unidentified {\em EGRET} error boxes performed with the Parkes radio telescope and the Australia Telescope Compact Array (ATCA), prior to the launch of \fermi.  The pulsar's period derivative $\dot P = 1.61 \times 10^{-14}$ s s$^{-1}$ yields a spin-down age of 92.1 kyr, a spin-down power $\dot{E}\sim 8.43 \times 10^{35}$ erg s$^{-1}$ and a dipolar magnetic field $B\sim 1.2 \times 10^{12}$ G. Thus, \psr\ may be considered a transition object between the young, Vela-like pulsars and the middle-aged ones.  The dispersion measure (DM=96.525 $\pm0.002$ cm$^{-3}$ pc) implies a distance of $\sim$2.3 kpc, according to the Galactic electron density model of  Cordes \& Lazio (2002).  Initially associated with the {\em EGRET} source 3EG\,1027$-$5817 upon its coincidence with the $\gamma$-ray error box, \psr\ was recently identified as a $\gamma$-ray pulsar by \fermi\   (Abdo et al.\ 2009a) and included both in the \fermi\ catalogue Bright Source List (Abdo et al.\ 2009b) and in the first \fermi-LAT catalogue of $\gamma$-ray pulsars (Abdo et al.\ 2010b).  Like for many \fermi\ pulsars, the $\gamma$-ray light curve is characterised by a double-peak profile, with a phase separation of $\sim 0.46$, which trails the single radio peak by $\sim 0.2$ in phase.  The pulsar's $\gamma$-ray spectrum is fit by a power-law (PL) with an exponential cut-off, with a photon index $\Gamma_{\gamma} = 1.25$ and a cut-off energy $E_{\rm c} =1.9$ GeV. Possible evidence for a PWN was also found in the \fermi\ data (Ackermann et al.\ 2011).  Recently, \psr\ was found to be spatially coincident with a TeV $\gamma$-ray source HESS\, J1023$-$575 (Abramowski et al.\ 2011), whose emission is probably associated with the PWN. In the X-rays, \psr\ was identified with a faint X-ray source detected by the \swift\ X-ray Telescope (XRT) in a short follow-up observation (Abdo et al.\ 2009a; Marelli et al.\ 2011) of the $\gamma$-ray source error box.  Unfortunately, the low detection significance ($\la 5 \sigma$) hampered the characterisation of the pulsar's X-ray spectrum.

While deeper X-ray observations  of PSR\, J1028$-$5819 are now in progress with \chan, no optical observations  have been performed, so far. Here, we present the first optical observations of \psr, performed with the \vltn\ (\vlt) as a part of a planned survey aimed at the optical identification of \fermi\ pulsars.
To complement our multi-wavelength analysis, we also present a re-analysis of the published \swift/XRT observation of the pulsar and the first results of a more recent, yet unpublished, {\em Suzaku} observation available in the public science data archive.

This paper  is organised as follows: observations,  data reduction and analysis are  described in  Sect. 2, while  results are  presented and discussed in Sect. 3 and 4, respectively.  Conclusions follow.

\section{Observations and data reduction}

\subsection{VLT observation}

PSR\, J1028$-$5819 was observed in service mode on February 10th and 11th, 2011 with the \vlt\ Antu telescope at the ESO Paranal    and the \forsn\  (\fors), a multi-mode camera for  imaging and long-slit/multi-object spectroscopy (Appenzeller  et  al.\ 1998).  \fors\  was used in imaging mode  and equipped with its standard MIT detector, a mosaic of two 2k$\times$4k CCDs optimised  for wavelengths  longer  than 6000  \AA.   With its  high-resolution colimator,   the  detector  has  a  pixel   scale  of  0\farcs125 (2$\times$2 binning), which  corresponds to a projected field--of--view (FOV) of 4$\farcm15  \times 4\farcm15$  over the CCD  mosaic. However,  due to vignetting, the effective sky coverage is smaller, and is larger for the upper CCD chip.   

Observations were performed with the standard low gain, fast read-out mode, with the high-resolution collimator,  and through the high-throughput  $b_{\rm HIGH}$ ($\lambda=4400$ \AA;  $\Delta \lambda=1035$\AA)  and $v_{\rm HIGH}$ ($\lambda=5570$ \AA;  $\Delta \lambda=1235$\AA) filters.  To allow for  cosmic ray removal, we obtained  sequences of short exposures (100 s) for a total integration time of 4800 s in both the  $b_{\rm HIGH}$ and $v_{\rm HIGH}$ filters.  We set such a short exposure time to minimise the saturation of a bright $V=13.12$ star close to the pulsar position, identified in the Digitised Sky Survey (DSS) images. Exposures  were taken  in  dark time  and under  photometric conditions, as   recorded    by   the    ESO    ambient   condition monitor\footnote{\texttt{http://archive.eso.org/asm/ambient-server}}, with an airmass of $\sim 1.2$ and image quality of $\sim 0\farcs8$, measured directly on the images by fitting the full-width at half  maximum (FWHM) of unsaturated field stars. 

Data were reduced through  the ESO \fors\ pipeline\footnote{\texttt{http://www.eso.org/sci/software/pipelines}} for  bias   subtraction and  flat--field   correction. Per each band, we aligned and average-stacked the reduced  science images using the  {\tt  swarp} program (Bertin et al.\ 2002), applying a $3  \sigma$ filter on  the single pixel average  to filter  out residual  hot and  cold pixels and cosmic ray hits.  Since all exposures were taken with sub-arcsec image quality, we did not apply any selection prior to the image stacking.  We applied the  photometric calibration by using  the  extinction-corrected night  zero points  computed by  the  \fors\ pipeline  and  available through  the instrument  data quality  control database\footnote{\texttt{www.eso.org/qc}}.  

\subsection{Swift observations}

\psr\ was pointed by the \swift/XRT on November 23rd 2008, during a Target of Opportunity (ToO) observation (Obs ID 00031298001) requested soon after the pulsar detection by {\em Fermi}.   The XRT (Burrows et al.\ 2005) is a grazing incidence Wolter I telescope equipped with an X-ray CCD imaging spectrometer, yielding a 110 cm$^2$ effective area at 1.5 keV,  a $23\farcm6 \times 23\farcm6$ FOV (2\farcs36/pixel), and an Half Energy Width of 15\arcsec\  in the 0.2--10 keV energy range.  The integration time was 9.6 ks (7.9 ks accounting for the exposure map correction). The observations was performed in PC mode. Data reduction and analysis were performed as in Marelli et al.\  (2011) but using the newest version (6.11) of the \texttt{HEAsoft} software package\footnote{\texttt{heasarc.gsfc.nasa.gov/docs/software/lheasoft/}}.
On the same visit, the \psr\ field was also observed in the near-UV with the Ultra-Violet Optical Telescope (UVOT) on board  \swift.  The UVOT (Roming et al.\ 2005) is  a 30 cm, f/12.7 Ritchey-Chr\'etien telescope  using a microchannel-intensified CCD detector which operates in photon counting mode, covering the $1600$--8000 \AA\ range over a $17\arcmin \times 17\arcmin$ FOV (0\farcs5/pixel).  Observations were performed with the UVM2 ($\lambda=2246$ \AA;  $\Delta \lambda=498$\AA) and the UVW2 ($\lambda=1928$ \AA;  $\Delta \lambda=657$\AA)  filters (Poole et al.\ 2008) for a total integration time of 4115 s and 5701 s, respectively. We retrieved the data  from the \swift\ science data archive and we combined the single exposures in order to obtain the best signal--to--noise (S/N). We used the UVOT data analysis software available in   \texttt{HEAsoft}
and the version 20110131 of the UVOT calibration files.

\subsection{Suzaku observations}

The field of PSR\,J1028--5819 was observed also by \emph{Suzaku} (Mitsuda et al.\ 2007) on July 22nd 2009. The observation (Obs ID 504045010) was performed with the pulsar located at the nominal aim-point position of the X-ray Imaging Spectrometer (XIS; Koyama et al.\  2007). The XIS consists of three front-illuminated (FI) CCD cameras (XIS0, XIS2 and XIS3), and one that is back-illuminated (BI; XIS1). Each CCD camera is combined with a single X-ray telescope (Serlemitsos et al.\ 2007) and has a single chip, yielding a 330 cm$^2$ (FI) or 370 cm$^2$ (BI) effective area at 1.5 keV, $18\arcmin\times18\arcmin$ field of view ($1\farcs05/$pixel), and $1\farcm8$ spatial resolution in the 0.2--12 keV energy range. One of the front-illuminated CCDs, XIS2, was not available at the time of our observation, since it suffered from a damage on 2006 November 9 and is unusable since then.  The observations was performed in full-frame mode.

We downloaded the data from the public \emph{Suzaku} science data archive through the DARTS interface\footnote{\texttt{www.darts.isas.jaxa.jp/astro/suzaku/}}. Data reduction and analysis of the XIS 
was performed with \texttt{HEAsoft} 
following the procedures described in the \emph{Suzaku} ABC Guide\footnote{\texttt{heasarc.gsfc.nasa.gov/docs/suzaku/analysis/abc/}} version 2. We used the task \texttt{xispi} to convert the unfiltered XIS event files to pulse-invariant channels and then \texttt{xisrepro} to obtain cleaned event files. For each XIS, $3\times3$ and $5\times5$ edit modes cleaned event data were combined using \texttt{xis5x5to3x3} and \texttt{xselect}. Following standard practices, we excluded times within 436 s of \emph{Suzaku} passing through the South Atlantic Anomaly and we also excluded the data when the line of sight was elevated above the Earths limb by less than $5^\circ$, or is less than $20^\circ$ from the bright-Earth terminator. Moreover, we excluded time windows during which the spacecraft was passing through the low cut-off rigidity of below 6 GV. Finally, we removed hot and flickering pixels using \texttt{sisclean}. The nominal on-source time is 23 ks. With all the aforementioned data selection criteria applied, the resulting total effective integration time is 20.2 ks. For the spectral analysis, the response matrices were generated  with \texttt{xisrmfgen} and using ray-tracing simulations with \texttt{xissimarfgen}.

\section{Results}

\subsection{VLT astrometry}

The  radio position of \psr\ has been obtained through interferometry observations with the ATCA  (Keith et al.\ 2008) and is: $\alpha =10^{\rm h}  28^{\rm m} 27\fs95$ ($\pm 0\fs1$) and  $\delta  = -58^\circ 19\arcmin 05\farcs22$ ($\pm 1\farcs5$), at MJD=54562. Due to the relatively small  difference of $\sim 3$ yrs between the epoch of the reference radio position and that of our \vlt\ observation, we can neglect any additional uncertainty on the pulsar position due to its unknown proper motion. If assuming the DM pulsar distance of 2.3 kpc  and a transverse velocity three times as large as the peak value of the radio pulsar velocity distribution ($\sim$ 400 km s$^{-1}$; Hobbs et al.\ 2005), the time difference would correspond to an additional uncertainty (in the radial direction) of $\sim$0\farcs3, much smaller than the nominal error on the radio position. No more recent (and more accurate) radio position for \psr\ has been obtained by the {\em Fermi} Pulsar Timing Consortium (Smith et al.\  2008), with Weltevrede et al.\ (2010) assuming the original ATCA position of Keith et al.\ (2008),  while the measured \swift/XRT position of the pulsar candidate counterpart,  $\alpha =10^{\rm h}  28^{\rm m} 27\fs69$ and  $\delta  = -58^\circ 19\arcmin 05\farcs16$ (Marelli et al.\ 2011), has an estimated radial uncertainty of $\sim 5\arcsec$. Thus, we assume the ATCA radio position as a reference. We note that {\em Simbad} reports a quite different position, $\alpha =10^{\rm h}  28^{\rm m} 39\fs8$ and  $\delta  = -58^\circ 17\arcmin 31\farcs0$, which is  still that given in the {\em Fermi}-LAT Bright Source List (Abdo et al.\ 2009b). 

To register the pulsar position on the \fors\ frames as precisely as possible, we re-computed their  astrometric solution which is, by default,  based on the coordinates of the guide star used for the telescope pointing.  As a reference, we used stars selected from the Guide Star Catalogue 2 (GSC-2; Lasker et al.\ 2008).  Although the recently released UCAC-3 catalogue (USNO CCD Astrograph Catalog; Zacharias et al.\ 2010) has a better astrometric accuracy then the GSC-2, it is cut at brighter magnitudes ($V\la 16$) so that most of the reference UCAC-3 stars are saturated even in the short-exposures \fors\ images. 
We selected GSC-2 stars evenly distributed in the field--of--view  but far from the CCD edges and the vignetted regions, where geometric distortions appear to be larger than expected. We noticed this effect for the first time  in our data set and, apparently, it affects all \fors\  images taken with the high-resolution collimator after 2009 (M. van den Ancker, priv. comm.). Its origin is still under investigation. We  measured the star centroids through Gaussian fitting  using the Graphical  Astronomy  and   Image  Analysis  ({\sc gaia})  tool\footnote{\texttt{star-www.dur.ac.uk/$\sim$pdraper/gaia/gaia.html}}  and used  the code  {\sc  astrom}\footnote{\texttt{www.starlink.rl.ac.uk/star/docs/sun5.htx/sun5.html}} to compute  the pixel-to-sky coordinate  transformation through an high-order polynomial, which accounts for the CCD distortions.   The  rms  of the astrometric  fits was $\sigma_{\rm r} \sim 0\farcs1$ in  the radial direction.   To this  value  we added  in  quadrature the  uncertainty $\sigma_{\rm tr}=0\farcs1$ on the registration of the \fors\ image on the  GSC2  reference frames, $\sigma_{\rm tr}= \sqrt{(3/N_{s})} \sigma_{\rm  GSC2}$,  where  $\sigma_{\rm  GSC2}=0\farcs3$ is  the  mean positional error  of the GSC2  coordinates  and $N_{\rm s}$ is  the number of stars used  to compute the astrometric solution (Lattanzi et al.\ 1997).  After accounting for the $\sim 0\farcs15$ accuracy of the link of the GSC2 coordinates to the International Celestial Reference Frame (ICRF), we thus estimate that  the overall ($1 \sigma$) uncertainty  of our \fors\  astrometry is $\delta_{\rm r}\sim0\farcs2$.  This is negligible compared with the absolute error on the reference pulsar position.

\begin{figure}
\centering
\includegraphics[height=8.5cm]{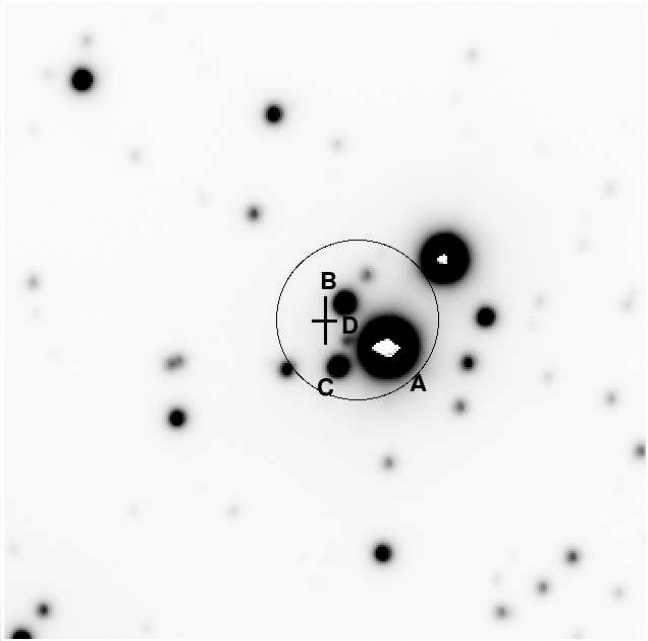}
  \caption{$40\arcsec \times 40\arcsec$ \vlt/\fors\ image of \psr\  in the $v_{\rm HIGH}$ filter (4800s).  North to the top, East to the left.   The cross marks the pulsar radio interferometry position (Keith et al.\ 2008), while the circle marks the position of the candidate X-ray counterpart detected by  \swift/XRT (Abdo et al.\ 2009a; Marelli et al.\ 2011). Stars close to the radio position are labelled (A-D). The bright Star A ($V=13.12$) is saturated in the \fors\ image. The intensity scale in the image has been stratched to visualise  the fainters stars (B-D).}
\end{figure}

\subsection{The search for the pulsar's optical counterpart}

The computed \psr\ position is shown in Fig.\ 1, overlaid on the \fors\ V-band image. As mentioned in Sectn. 2.1, it falls $\la 4\arcsec$ from a bright ($V=13.12$) field star (Star A) west of it. Unfortunately,  since its halo extends to $\la 1\arcsec$ from the pulsar position we could not mask Star A without risking of severely hampering our observations.  Indeed,  since the angular separation between Star A's halo and the pulsar is compatible with the telescope pointing accuracy and smaller than the absolute uncertainty on the pulsar's position (see above), this could have been accidentally covered by one of the \fors\ occulting bars ($\sim 12\farcs5$ width with the high-resolution collimator). Moreover, Star A is part of a triplet (Star B, C) of relatively bright stars which was not resolved in the DSS-2 images, the only ones available for a  quick look of the pulsar field, due to their coarse spatial resolution. This would have caused the mask to be put even closer to the pulsar position, thus increasing the risk of an accidental occultation of our target. We note that the northern star of the triplet (Star B) overlaps the pulsar position. However, its relatively high flux ($B\sim20.3$; $V\sim17.5$) obviously rules it out as a possible candidate counterpart to the pulsar (e.g., Mignani 2011). The same argument also applies to the southern star of the triplet (Star C) which is of comparable brightness  ($B\sim20.7$; $V\sim17.5$).  

A fourth, fainter star (Star D) is visible in the V-band image at $\sim 2\arcsec$ from the nominal radio pulsar position, superimposed to the halo of the much brighter Star A, while in the B-band image it can not be clearly resolved, partially because of the slightly worse image quality. The offset is comparable with the $1 \sigma$ uncertainty on the pulsar position, which, in principle, makes it a possible candidate counterpart. We computed the flux of Star D  through customised aperture photometry using a small aperture of 4 pixel radius (0\farcs5) to minimise the contribution of the bright Star A's  halo and optimise the S/N. We sampled the local background within circular areas of the same radius as the photometry aperture placed below and above the Star D's position. We then computed the aperture correction using the growth curves of a few unsaturated field stars.  The computed magnitude is $V\sim 20.5 \pm 0.2$, where the photometric error is dominated by the uncertainty on the subtraction of the background of the bright Star A's halo, which makes it also an improbable candidate pulsar counterpart.

No other object is  clearly visible at, or close to, the radio pulsar position. Although our images were taken under sub-arcsec seeing conditions and with short exposure times to minimise the flux contamination of Star A, its bright halo still significantly affects the background close to the pulsar position, hampering the detection of its faint counterpart. We tried to subtract Star A using the model image point spread function (PSF), computed by fitting the intensity profile of several unsaturated field stars. To this aim,  we used the tools {\tt daofind}, {\tt nstar}, {\tt peak} and {\tt substar} available in the {\sc IRAF} package {\tt daophot}. Since Star A is partially saturated, this procedure obviously leaves some residuals at the PSF core, but can account for the subtraction of most of the PSF and its wings.  Similarly, we also PSF-subtracted stars B, C, and D.

\begin{figure*}
\centering
\includegraphics[height=7.5cm,angle=0]{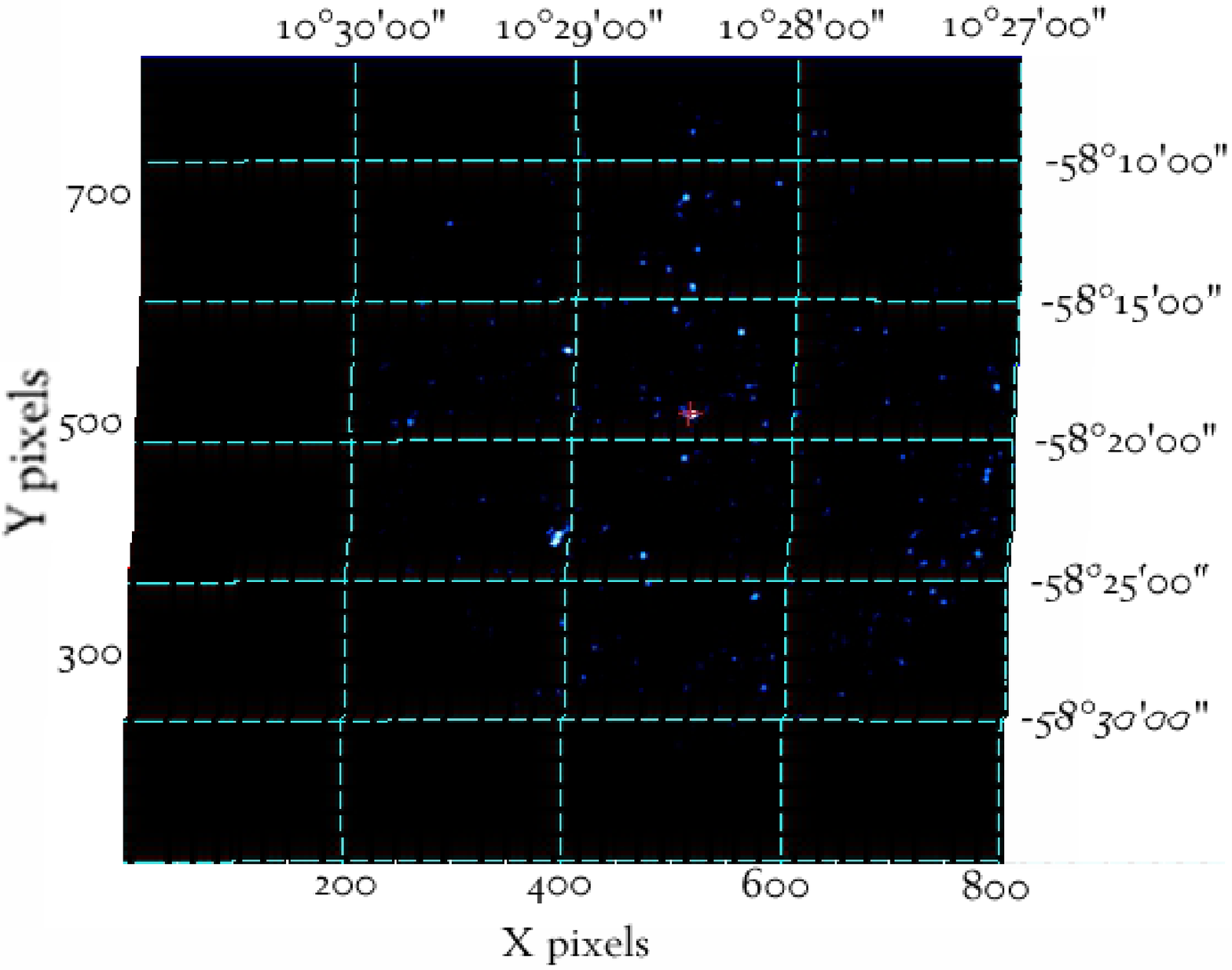}
\includegraphics[height=7.5cm,angle=0]{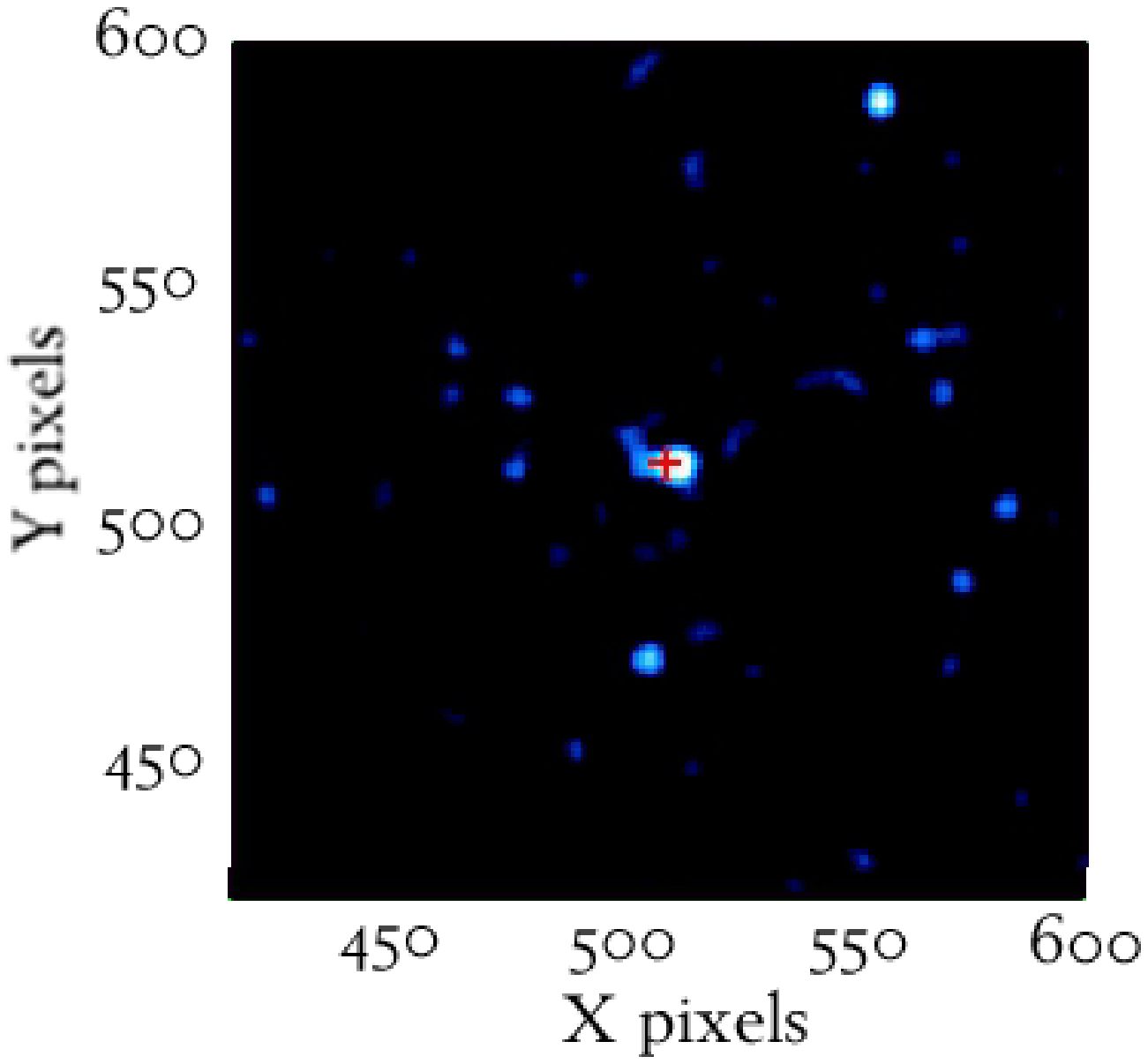}
\caption{\emph{Left panel:} \swift/XRT image of the field of  PSR\,J1028--5819 in the 0.3--10 keV energy range. The source highlighted by the red circle is the pulsar's X-ray counterpart (Abdo et al.\ 2009a; Marelli et al.\ 2011). \emph{Right panel: } Zoom-in on the source position. The red cross marks the ATCA position of PSR\,J1028--5819 (Keith et al.\ 2008).}
\label{xrt-fov}
\end{figure*}

We did not detect any excess signal  in the PSF-subtracted images, which can be associated with emission from a point-like source at the pulsar position. Thus, we concluded that \psr\ was not detected in our \vlt\ images. We computed the optical flux upper limits in the PSF-subtracted B and V bands images from the rms of the background sampled in the pulsar radio error box within cells of $7\times7$ pixels,  i.e. of size equal to the computed image FWHM, and we applied the computed aperture correction. We then derived $3 \sigma$ upper limits of $V\sim 25.3$ and $B\sim 25.4$. Unfortunately, these limits are affected by the residuals of the PSF subtraction of the four stars, with Star A and B contributing the most, which increase the local sky background by a factor of  4 with respect to other regions in the field, where the limiting magnitudes are about 1.5 magnitudes deeper.

\subsection{The pulsar's X-ray counterpart: \swift\ observation}

It is interesting to compare our derived optical flux upper limits on \psr, the deepest obtained so far, with the extrapolation of the best-fit to the X-ray spectrum of the pulsar's counterpart.  Unfortunately, this is poorly constrained by the available \swift/XRT observation (Fig.\ 2), due to the short integration time (7.9 ks) and the low count statistics. We extracted all counts within a 20\arcsec radius using a similar sized background region. This yields $(2.41\pm0.69) \times 10^{-3}$  counts s$^{-1}$ in the 0.3--10 keV energy range.   We re-analysed the XRT spectral data using the newest version (12.7) of the {\sc xspec} fitting package (Arnaud et al.\ 1996). As in Marelli et al.\  (2011), we found a satisfactory fit with  a PL with fixed photon index $\Gamma_{\rm X}=2$, assuming a fixed hydrogen column density  $N_{\mathrm{H}}\sim 5\times10^{21}$ cm$^{-2}$ inferred from the Galactic value along the line of sight using the  \texttt{Heasarc} {\tt nH} tool\footnote{http://heasarc.gsfc.nasa.gov/cgi-bin/Tools/w3nh/w3nh.pl} scaled for the pulsar distance of 2.3 kpc assuming a constant distribution.  We note that this value of the $N_{\mathrm{H}}$ is about a factor of $\sim 3$ smaller than assumed by Abdo et al.\ (2009a), $N_{\mathrm{H}} = 1.59\times10^{22}$ cm$^{-2}$, whose determination, however, is not explained.  For these spectral parameters, the \swift/XRT spectrum corresponds to an unabsorbed 0.5--10 keV X-ray flux $F_{\rm X} = (1.5\pm0.5) \times 10^{-13}$ erg cm$^{-2}$ s$^{-1}$.

\subsection{The pulsar's X-ray identification revisited}

The uncertainty on the pulsar's X-ray spectrum obviously affects the comparison with our optical flux measurements. Moreover, the positional coincidence between the  \swift/XRT error circle of the pulsar candidate counterpart and Star A  (Fig.\ 1) suggests that, in principle, this might contribute to the observed X-ray emission, which would make the characterisation of the pulsar X-ray spectrum even more uncertain. 
As a most extreme possibility, it cannot be  ruled out that the \swift/XRT source  is, actually,  the X-ray counterpart to Star A and not to the pulsar.  
The star was clearly detected by the \swift/UVOT (see Fig.\ 3) but, admittedly,  its possible contribution to the emission of the X-ray source was not accounted for by Marelli et al. \ (2011).  Thus, we tried to evaluate such contribution by determining the star's spectral type via multi-band photometry. 

For the \swift/UVOT observations, we used the {\sc ftool}  \texttt{uvotsource} to determine the count rate of the star in both the UVM2 and in the UVW2  filters. We used a 2\arcsec  aperture in order to minimise  contamination from the nearby star north-west of it.      The count rate was aperture-corrected to a radius of 5\arcsec\  in order to apply the count-rate to magnitude conversion using the zero points from the the \swift/UVOT calibration (Breeveld et al.\ 2010, 2011). The computed magnitudes (in the Vega system) are $m_{UVM2} =  16.29 \pm 0.10$ and $m_{UVW2} =  16.51 \pm 0.11$, where the associated errors are purely statistic (systematic errors are $\sim 0.03$ magnitudes). 
The star's  optical magnitudes are $R_{\rm F} =12.87$ and $V=13.12$, according to the GSC-2 (Lasker et al.\ 2008), 
while the near-infrared magnitudes, as derived from 2MASS (Skrutskie et al.\ 2006), are $J=12.38$; $H=12.16$; $K_s=12.09$. 
Unfortunately, the UVOT, DSS, and 2MASS, images have not enough angular resolution to resolve the triplet of stars around Star A, while the star $\approx 7\arcsec$ north-west of it (Fig.\ 1) is clearly resolved in all of them (see, e.g.  Fig.\ 3). Thus, the above magnitudes must be taken with due care. However, from our \vlt\ data we see that Star B and C are at least a factor of $\sim$ 50 fainter than Star A in both the B and V bands.  Thus, their contribution only marginally affects the estimate of the star colours. 

\begin{figure}
\centering
\includegraphics[height=8.5cm]{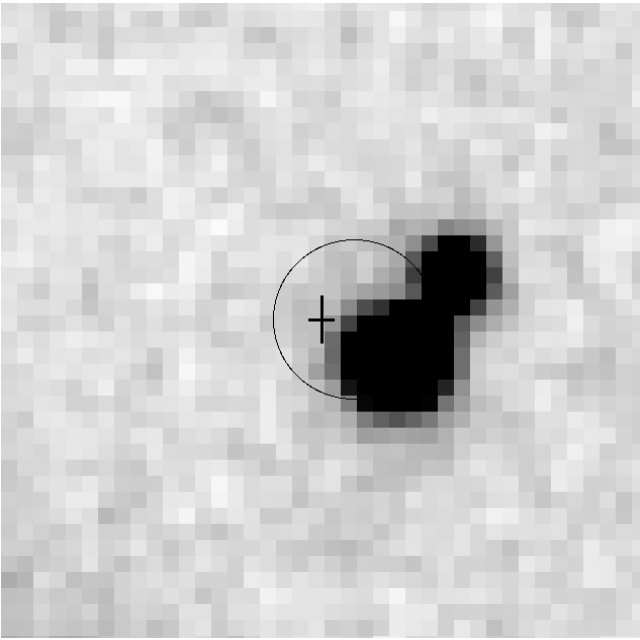}
  \caption{Same field as in Fig.1 but obtained by the \swift/UVOT through the UVM2 filter (4115 s). The group of stars around Star A (see Fig. 1) is unresolved at the UVOT spatial resolution.}
\end{figure}

\begin{table*}
\centering
\begin{minipage}{13.9cm}
\caption{Summary of the \emph{Suzaku}/XIS spectral analysis. Errors are at a 1$\sigma$ confidence level for a single parameter of interest.}
\label{xis-fits}
\begin{tabular}{@{}ccccccc}
\hline
\hline
Model & $N_{\mathrm{H}}$ & $\Gamma_{\rm X}$ & $kT$ & \multicolumn{2}{c}{Observed / unabsorbed flux$^{a}$} & $\chi^2_\nu$/dof\\
 & ($10^{21}$ cm$^{-2}$)&  & (keV) & \multicolumn{2}{c}{($10^{-13}$ erg cm$^{-2}$ s$^{-1}$)} & \\
\hline
Power law & $<$1 & $1.7\pm0.2$ & -- & $1.5^{+0.1}_{-0.4}$ & $1.5\pm0.2$ & 1.09/21\\
Bremmstrahlung & $<$0.5 & -- & $8^{+12}_{-3}$ & $1.4^{+0.1}_{-0.5}$ & $1.4\pm0.2$ & 1.16/21 \\
Raymond-Smith & $<$0.6 & -- & $7^{+10}_{-3}$ & $1.4^{+0.1}_{-0.4}$ & $1.5^{+0.1}_{-0.3}$ & 1.17/21 \\
Blackbody & $<$0.5 & -- & $0.5\pm0.1$ & $0.7\pm0.2$ & $0.7\pm0.1$ & 1.97/21 \\
\hline
\end{tabular}
\begin{list}{}{}
\item[$^{a}$] In the 0.5--10 keV energy range.
\end{list}
\end{minipage}
\end{table*}

The 
colours of Star A are compatible with those of an early F-type main sequence star, assuming an interstellar extinction corresponding to an $E(B-V)\sim 0.09$.  
According to the relation of Predhel \& Schmitt\ (1995), this value would correspond to an $N_{\mathrm{H}} \sim0.5 \times10^{21}$ cm$^{-2}$, i.e. lower than estimated 
from scaling  the Galactic value along the line of sight for the \psr\ distance,
suggesting that Star A is, possibly, foreground to the pulsar, consistent with a distance of $\approx 1$ kpc derived from its distance modulus. Usually, F-type stars have an X-ray--to--optical flux ratio $F_{\rm X}/F_{\rm V} \approx 10^{-3}$--$10^{-2}$ (e.g. Krautter et al.\ 1999) which might yield an X-ray flux comparable with that of the \swift/XRT source.  Thus, it is possible that this is, indeed,  the X-ray counterpart to Star A and not to the pulsar. However, any conclusion still depends  on a better characterisation of the X-ray source spectrum and flux. To this aim, we have used the archival {\em Suzaku} observation (see Sectn. 2.3).

\subsection{The pulsar's X-ray counterpart: Suzaku observation}

\begin{figure*}
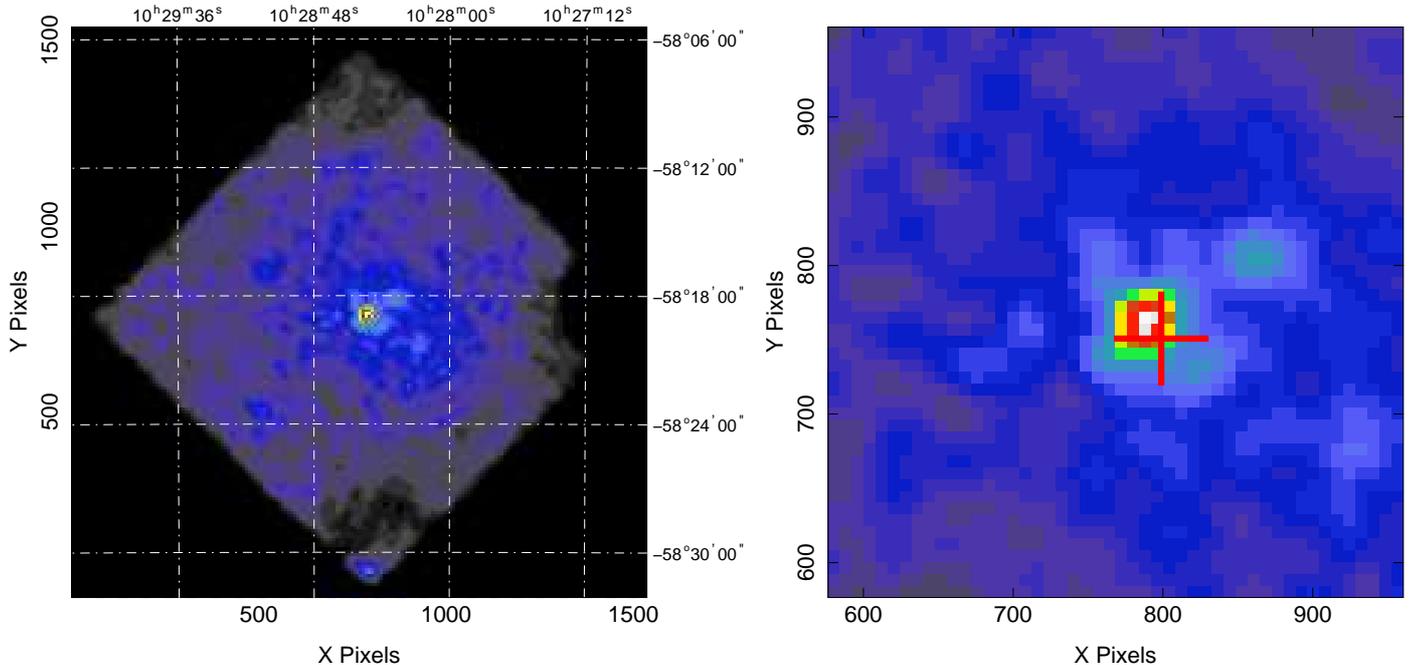

\centering
\resizebox{\hsize}{!}{
\includegraphics[angle=0]{18869_f5.ps}
\includegraphics[angle=0]{18869_f6.ps}}
\caption{\emph{Left panel:} \emph{Suzaku}/XIS image of the field of  PSR\,J1028--5819 in the 0.2--10 keV energy range.  The pulsar's X-ray counterpart is at the centre of the field--of--view.
\emph{Right panel: } Zoom-in on the source position. Like in Fig. 2, the red cross marks the ATCA position of PSR\,J1028--5819 (Keith et al.\ 2008).}
\label{xis-fov}
\end{figure*}

The \emph{Suzaku} image of the PSR\,J1028--5819 field obtained combining the data from the three XIS cameras is shown in Fig.~\ref{xis-fov}. Using the sliding-cells source detection algorithm in the \texttt{ximage} package, a source is detected at the centre of the XIS FOV with a signal-to-noise ratio of 18 in the 0.2--10 keV energy range. The source best-fit position, $\alpha = 10^h28^m 28\fs7$, $\delta =-58\degr18\arcmin 59\farcs3$ (J2000), is consistent, within the absolute astrometry accuracy of \emph{Suzaku} ($\sim$$20\arcsec$; Uchiyama et al.\ 2008), with the ATCA coordinates of PSR\, J1028--5819 and with those of the source detected by \swift.

We extracted the source counts from each XIS detector using a circular region of $0\farcm7$ radius centred on the best fit \texttt{ximage} position. This comparatively small region, corresponding to $\sim$50\% of the encircled energy fraction,  was chosen in order to improve the signal--to--noise ratio of the spectra and reduce the possible contamination from the diffuse emission apparent in Fig.~\ref{xis-fov}. The background was extracted from an annulus with radii 3\arcmin\ and 5\arcmin\ centred on the X-ray source. Accounting for the counts of all the XIS detectors, the 0.5--10 keV net source count rate is $(2.7\pm0.3)\times10^{-3}$ counts s$^{-1}$, which is not a very high statistics for a detailed spectral analysis. To increase photon statistics, we merged the source spectral files and response matrices using \texttt{addascaspec}. The 3-XIS spectrum was binned so as that each bin contains a minimum of 15 counts.
As done for the {\em Swift}/XRT spectrum, we performed our spectral fits using the \texttt{XSPEC} fitting package (version 12.7). The fit with an absorbed PL model  (Fig.\ 5) yields a spectral index $\Gamma_{\rm X} = 1.7\pm0.2$ and an absorption poorly constrained at $N_{\mathrm{H}}<10^{21}$ cm$^{-2}$ at (1$\sigma$), with a reduced $\chi^2_\nu = 1.09$ for 21 degrees of freedom.  This corresponds to an unabsorbed 0.5--10 keV X-ray flux $F_{\rm X} = (1.5 \pm 0.2) \times 10^{-13}$ erg cm$^{-2}$ s$^{-1}$, very well consistent with the value derived by Marelli et al.\ (2011) from the \swift/XRT observation in a similar energy range. We note that our upper limit on the $N_{\mathrm{H}}$ is lower than the value inferred by scaling the Galactic value along the line of sight (Marelli et al.\ 2011), although it is not incompatible with the latter once the associated uncertainties are taken into account. We also tried other single-component models, like a blackbody (BB), a Bremsstrahlung, and a Raymond-Smith plasma (Raymond \& Smith 1977). However, they all give higher $\chi^2$ values than the PL,  although the fits obtained using the Bremsstrahlung and Raymond-Smith plasma models are only slightly worse than obtained using the PL (see Table~\ref{xis-fits} for a summary). To improve our results, we tried to fit the joint \swift/XRT and {\em Suzaku}/XIS spectra, but due to the low statistics of the former observation our results did not change significantly. 

Our spectral fits  confirm the qualitative results of Marelli et al.\ (2011), based on a lower statistics. In particular, the good fit with a PL ($\chi^2_\nu = 1.09$)
suggests that the spectrum of the X-ray source is non-thermal. Thus, it would not be  produced by coronal emission from a main sequence star, like Star A.  This would confirm that the X-ray source detected by both the \swift/XRT  and {\em Suzaku} is, indeed, the pulsar's X-ray counterpart. Due to the still limited statistics, we could not try spectral fits with two-component models like, e.g. a PL+BB.  Thus, we can not rule out that Star A might, indeed, contribute at some level to the observed X-ray emission. Deeper observations with  \xmm\ will allow one to obtain a more precise characterisation of the X-ray spectrum of the \psr\ counterpart, while optical spectroscopy of Star A will provide a far more precise classification of its spectral type, allowing one to better quantify its expected X-ray flux. In addition, high-resolution \chan\ images will allow one to spatially resolve a possible X-ray source associated with Star A from the pulsar's X-ray counterpart and secure its identification through the detection of X-ray pulsations\footnote{We note that the frame times of the \swift/XRT in PC mode and of the {\em Suzaku}/XIS in full-frame mode are 2.507 s and 8s, respectively, which makes it impossible to seach for pulsations at the the \psr\ 91 ms period}.

We also inspected the \emph{Suzaku} data to search for possible evidence of a PWN associated with \psr. As mentioned above,  diffuse emission on a scale of a few arcminutes isobserved around the pulsar's  X-ray counterpart (Fig.\ 4), with an apparent elongation in the south-west direction.  
The count-rate in this region is $(5.3\pm 0.5)\times10^{-4}$ counts s$^{-1}$ arcmin$^{-2}$ in the 0.2--10 keV,
computed after summing the counts collected with the three XIS detectors. 
However, since each XRT/XIS has a different position-dependent and broad 
PSF, with a complicated (roughly cross-shaped) and asymmetric profile (see, e.g. Sugizaki et al.\ 2009), it is difficult to determine whether the diffuse emission observed around the pulsar's X-ray counterpart is real or not. 
As a test, we compared its surface brightness profile  with that of a reference X-ray source.  Since there are no suitable sources detected in the \psr\ field--of--view (Fig. 4), we used as a reference an X-ray source detected in another {\em Suzaku} observation (Obs ID 804017020) and identified with the ms-pulsar PSR\, J1231$-$1411. Since both sources were observed on-axis, our comparison is not affected by the position dependance of the XIS PSF.  In both cases, we selected photons in the 0.3--10 keV energy range in all the three XIS detectors and we used the {\tt xissim} tool to simulate and subtract the background, accounting for the detector vignetting. Then, we extracted the radial intensity profiles, averaged over a 360$^{\circ}$ angle, for both the \psr\ counterpart and the reference source.  Finally, we normalised the counts of the reference source, computed within a radial distance bin of $10\arcsec$, to that of the \psr\ counterpart.  As seen from Fig.\ 6, the latter seems to feature an excess of counts with respect to the reference X-ray source at radial distances $\sim 80\arcsec$--$200\arcsec$, although such excess is much less evident at larger distances. This, coupled with the low count-rate of the diffuse emission, makes it difficult to determine whether it is real, hence possibly associated with a PWN or with an unresolved field source,  and to quantify its contribution to the observed emission of the \psr\ X-ray counterpart, in particular at angular distances smaller than $80\arcsec$.    \chan\ observations will be crucial to confirm the presence of the diffuse emission observed by {\em Suzaku}. If real, future measurements of the \psr\ radio proper motion will, then, allow one to search for  a possible alignment with the semi-major axis of the candidate PWN. 
 
\begin{figure}
\centering
\includegraphics[height=8.5cm,angle=270,clip=]{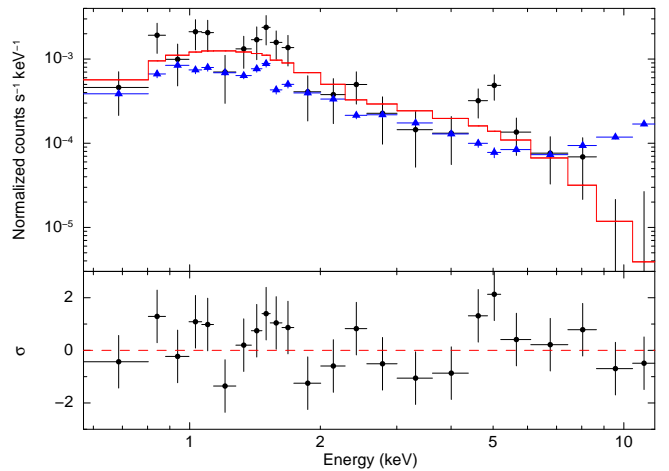}
  \caption{{\em Suzaku}/XIS spectrum of the candidate \psr\  X-ray counterpart obtained using data from the three XIS CCDs. The background level is marked in blue and the best-fit PL model in red.}
\end{figure}

\begin{figure}
\centering
\includegraphics[height=8.5cm,angle=270,clip=]{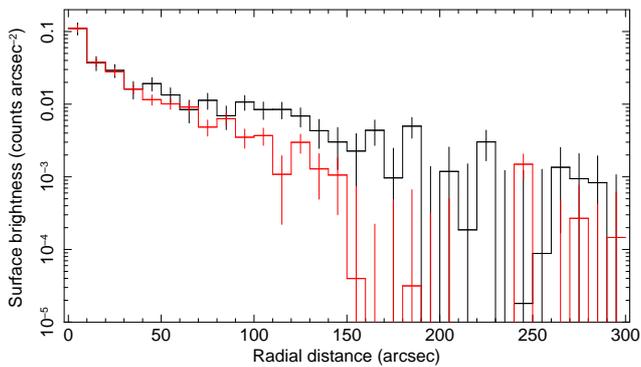}
  \caption{Surface brightness profiles of the \psr\ X-ray counterpart (black) detected in the {\em Suzaku}/XIS data and of a reference X-ray source  (red). The excess with respect to the reference source is different from zero at the $\sim5 \sigma$ level, (accounting for statistical errors only). This suggests the presence of diffuse X-ray emission surrounding \psr\ on angular scales of a few arcminutes.}
\end{figure}

\section{Discussion}

\subsection{The pulsar luminosity}

We used our magnitude upper limits in the B-band\footnote{For comparison with the luminosities quoted in Zharikov et al.\ (2006), which are computed in the B band}to compute the corresponding extinction-corrected optical flux and luminosity of \psr. As a reference for the interstellar extinction correction we assumed the upper limit on the hydrogen column density ($N_{\mathrm{H}}=10^{21}$ cm$^{-2}$) derived from the fits to the {\em Suzaku} X-ray spectrum. According to the relation of Predhel \& Schmitt\ (1995), this corresponds to an interstellar extinction  $A_{\rm V} \sim 0.55$.  Using the extinction coefficients of Fitzpatrick\ (1999), the extinction-corrected B-band upper limit is, thus, $F_{\rm B} \la 9 \times 10^{-16}$ erg cm$^{-2}$ s$^{-1}$.
If we assume for the pulsar the unabsorbed {\em Suzaku} X-ray flux derived from the best-fit PL model, $F_{\rm X} = (1.5 \pm 0.2) \times 10^{-13}$ erg cm$^{-2}$ s$^{-1}$, 
our extinction-corrected B-band upper limit 
would correspond to  an X-ray--to--optical flux ratio $F_{\rm X}/F_{\rm B} \ga 150$,  too low to independently suggest that the X-ray source detected by {\em Swift} and {\em Suzaku} is indeed the \psr\ counterpart.  
We computed the upper limit on the pulsar optical luminosity using as a reference its DM-based distance of $\sim$2.3 kpc (Keith et al.\ 2008). We note that while the uncertainty on the measured DM is negligible (DM=96.525 $\pm0.002$ cm$^{-3}$ pc), the  Galactic electron density model of  Cordes \& Lazio (2002) is known to have an average uncertainty of $\sim$30\% which, however, can be much larger for individual objects.
As in the first \fermi-LAT catalogue of $\gamma$-ray pulsars  (Abdo et al.\ 2010a), we assumed the typical $30\%$ uncertainty on the nominal DM-based distance
of \psr. This corresponds to an uncertainty of $0.7$ kpc, which we assume in the following discussion.
The B-band flux upper limit corresponds to an optical luminosity 
$L_{\rm B} \la 9.2 \times 10^{29}$ erg s$^{-1}$ for \psr, after accounting for the assumed distance uncertainty.
 Given the pulsar spin-down age of 92.1 kyr, we can assume that its, yet undetected, optical emission be purely non-thermal, as normally observed in the youngest pulsars  (see, e.g. Mignani 2011).  Thus, for the pulsar's spin-down power, $\dot{E}\sim 8.43 \times 10^{35}$ erg s$^{-1}$, our luminosity upper limit implies an optical emission efficiency 
$\eta_{\rm B} \equiv L_{\rm B}/\dot{E} \la 1.09 \times 10^{-6}$. Both limits are consistent with the optical luminosity and efficiency computed for most pulsars older than $\sim $0.1 Myr (Zharikov et al.\ 2006) and confirm an apparent turnover in the pulsar optical emission properties.
Similarly, 
the pulsar X-ray flux derived from the best-fit PL mode would correspond to an X-ray luminosity 
$L_{\rm X} = (9.5\pm 5.9) \times 10^{31}$ erg s$^{-1}$, after accounting for the X-ray flux and pulsar distance uncertainties.  
 This  value would imply an X-ray emission efficiency 
$\eta_{\rm X} \equiv L_{\rm X}/\dot{E} = (1.13 \pm 0.70) \times 10^{-4}$, for an assumed beaming factor $f=1$. The derived X-ray efficiency would be consistent, within the large scatter, with the trend observed for the {\em Fermi} pulsars (Marelli et al.\ 2011).

\subsection{The pulsar multi-wavelength spectrum}

We compared the extinction-corrected B and V-band spectral flux upper limits with the extrapolations in the optical domain  of the X-ray and $\gamma$-ray spectra of \psr.  For the X-ray spectrum we used as a reference our best-fit PL model with spectral index $\Gamma_{\rm X}= 1.7\pm0.2$, while for the $\gamma$-ray spectrum we assumed a PL with exponential cutoff, with spectral index $\Gamma_{\gamma} = 1.25\pm0.17$ and cutoff energy $E_{\rm c} = 1.9 \pm 0.5$ GeV, obtained from a re-analysis of the {\em Fermi} data (see also Marelli et al.\ 2011). The multi-wavelength spectral energy distribution (SED)  of \psr\  is shown in Fig.\ 7, where we accounted for the $1 \sigma$ uncertainty on the extrapolations of the X and $\gamma$-ray PL spectra.  
The optical spectral flux upper limits are clearly above the extrapolation of the $\gamma$-ray PL, which does not rule out that the optical emission of \psr\ can be associated with the low-energy tail of the $\gamma$-ray spectrum, as suggested for most $\gamma$-ray pulsars identified in the optical (Durant et al.\ 2011). This would imply a break in the optical--to--X-ray spectrum,  as observed in most pulsars (see, e.g. Mignani et al.\ 2010). In this case, the observed optical flux would be $B=28$--34, a range only marginally reachable by current 8m-class telescopes. On the other hand, the optical upper limits are also compatible with the extrapolation of the X-ray PL, which does not allow yet one to confirm the presence of a  possible spectral break in the optical--to--X-ray spectrum.  In general, the SED suggests that no single PL can fit the optical--to--$\gamma$-ray spectrum of the pulsar, with at least one spectral break required between the X and $\gamma$-ray energy ranges. This trend has been observed also in other {\em Fermi}  pulsars (e.g., Mignani et al.\ 2011), probably related to a complex particle energy and density distribution  in the neutron star magnetosphere.

\begin{figure}[h]
\centering
\includegraphics[height=7cm,angle=0,clip=]{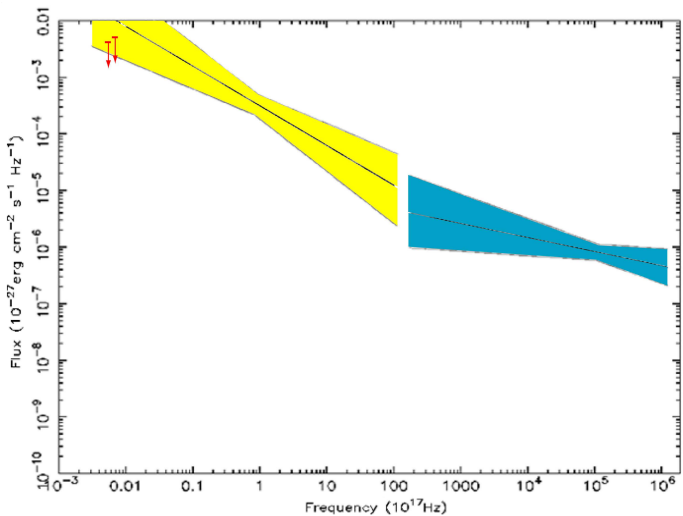}
\caption{Upper limits on the extinction-corrected optical fluxes of \psr\ compared with the low-energy extrapolations of the X-ray (solid black) and $\gamma$-ray (solid blue) spectral models that best fit the {\em Suzaku}  and {\em Fermi} (Abdo et al.\ 2009a) data.  The plotted optical flux upper limits are corrected for interstellar extinction based upon an $N_{\mathrm{H}}=10^{21}$ cm$^{-2}$. The dotted  line corresponds to the PL  X-ray spectrum. The yellow and blue-shaded areas  indicate the $1 \sigma$ uncertainty on the extrapolations of the X and $\gamma$-ray PLs, respectively.  }
 \end{figure}
 
\subsection{The candidate PWN}

Finally, we verified whether the diffuse emission marginally detected in the {\em Suzaku} data (Sectn. 3.5) would be compatible, if it were indeed a PWN, with the \psr\ energetics and distance. PWNe have been detected in the X-rays around a number of young and energetic pulsars (see, e.g. Kargaltsev \& Pavlov 2010 for a recent compilation), with luminosities  equal to $\sim 10^{-4}$--$10^{-2}$ of the pulsar spin-down power $\dot{E}$.  In the case of \psr,  one would then expect an X-ray luminosity  $L_{\rm X, PWN} \sim 8.5 \times 10^{31}$ erg s$^{-1}$ for its PWN, assuming $L_{\rm X, PWN}/\dot{E}=10^{-4}$. At the \psr\ distance of $\sim 2.3\pm0.7$ kpc, this would correspond to an integrated flux $F_{\rm X, PWN} \sim (1.3 \pm 0.8)  \times 10^{-13}$ erg cm$^{-2}$ s$^{-1}$ for the PWN.  By using the {\tt WebPIMMS} tool\footnote{http://heasarc.nasa.gov/Tools/w3pimms.html} assuming a PWN spectral index $\Gamma_{\rm X,PWN} =2$ and  an hydrogen column density $N_{\mathrm{H}} \sim 10^{21}$ cm$^{-2}$, this would correspond to a count-rate of  $(8\pm5) \times 10^{-3}$ counts s$^{-1}$ in the 0.2--10 keV energy range. Following, e.g.  Bamba et al.\ (2010), a PWN around a $\sim 100$ kyr old pulsar would have an angular extent of $\sim 11\farcm5$--$21\farcm5$ at a distance of $2.3\pm0.7$ kpc, i.e.  comparable to  that  of the diffuse emission detected in the {\em Suzaku} data.   For the expected PWN extension, we then derive an expected count-rate of $\sim(0.2-3) \times 10^{-5}$ counts s$^{-1}$ arcmin$^{-2}$.  This value is up to a factor of 30  lower than the actual count-rate measured for the diffuse emission  around \psr. However, it is not incompatible with it if one assumes, e.g. either a correspondingly larger $L_{\rm X, PWN}/\dot{E}$, which would still be within the observed range for the known PWNe, or a non-uniform PWN surface brightness.  Thus, it is indeed plausible that the diffuse emission detected by {\em Suzaku} might be associated with the \psr\ PWN.

\section{Summary and conclusions}

We performed the first deep observations of the {\em Fermi} pulsar \psr\ with the \vlt. Unfortunately, due to the presence of a bright star F-type star (Star A; $V=13.12$) $\sim 4\arcsec$ away from the pulsar radio position (Keith et al.\ 2008) we could only set limits of $V\sim25.3$ and $B\sim 25.4$ on its optical flux, admittedly affected by the residuals after the subtraction of Star A's PSF and by the coarse accuracy of the radio coordinates. These limits correspond to an optical luminosity $L_{\rm B} \la 9.2 \times 10^{29}$ erg s$^{-1}$ and emission efficiency $\eta_{\rm B} \la 1.09 \times 10^{-6}$, consistent with the pulsar spin-down age of 92.1 kyr.   
We also analysed archival {\em Suzaku} observations of \psr, which provide a better detection of the pulsar's X-ray candidate counterpart and support its identification  as a pulsar from a better characterisation of the X-ray spectrum, which is best-fit by a PL with spectral index $\Gamma_{\rm X} = 1.7\pm0.2$. The inferred pulsar's X-ray efficiency ($\eta_{\rm X} \sim 1.13 \times 10^{-4}$) would be consistent with the trend observed for the {\em Fermi} pulsars (Marelli et al.\ 2011).  The pulsar multi-wavelength spectrum cannot be fit by single PL, implying  at least one spectral break in the optical--to--$\gamma$-ray energy range, as observed in other {\em Fermi} pulsars. We also find a possible evidence for the presence of X-ray diffuse emission around \psr, with a surface brightness and angular extent consistent with the observed X-ray luminosity and size of known PWNe. 
More observations are required to complete the multi-wavelength study of \psr\ at low energies. On the optical side, high-spatial resolution observations with the \hst, possibly in the near-UV to exploit the low interstellar extinction along the line of sight, coupled with a much more accurate radio position, are needed to resolve the pulsar emission against the PSF wings of Star A. On the X-ray side, follow-up observations with \xmm\ and \chan\  are needed to obtain a better characterisation of the pulsar's X-ray spectrum, accounting for any possible contribution from Star A, and detect X-ray pulsations, which will confirm the proposed identification. \chan\  observations will also be crucial in confirming the existence of the diffuse emission marginally detected by {\em Suzaku}  and study its spectrum and morphology.  

\begin{acknowledgements}
A key contributor to the \fors\ pipeline, Carlo Izzo (ESO), prematurely passed on June 23 2011 after fighting courageously a short illness. Thanks to Carlo's work, the results from many  \fors\ observations were published, making his inheritance forever lasting. RPM thanks Paul Ray for checking the \psr\  coordinates and Mario van den Ancker and Sabine Moheler (ESO) for investigating the \fors\ geometric distortions. The authors thank Fabio Mattana for useful comments. This research made use of data obtained from Data ARchives and Transmission System (DARTS), provided by Center for Science-satellite Operation and Data Archives (C-SODA) at ISAS/JAXA. PE acknowledges financial support from the Autonomous Region of Sardinia through a research grant under the program PO Sardegna FSE 2007--2013, L.R. 7/2007 ``Promoting scientific research and innovation technology in Sardinia''. This research has made use of data obtained from the High Energy Astrophysics 
Science Archive Research Center (HEASARC) and the Leicester Database and Archive Service 
(LEDAS), provided by NASA's Goddard Space Flight Center and the Department of Physics and Astronomy, Leicester University, UK, respectively. 
SRO acknowledges the support of the UK Space Agency.
\end{acknowledgements}

\end{document}